\documentclass[amssymb,amsmath,twocolumn,superscriptaddress,nofootinbib,floatfix]{revtex4-2}

\usepackage{graphicx}
\usepackage{bm}
\usepackage{amssymb,amsmath}
\usepackage{mathrsfs}
\usepackage{latexsym}
\usepackage{color}
\usepackage[normalem]{ulem}
\usepackage{dcolumn}

\usepackage[colorlinks=true,citecolor=blue,urlcolor=blue]{hyperref}
\usepackage[usenames,dvipsnames]{xcolor}
\usepackage{xspace}
\usepackage{graphicx}
\usepackage{multirow}
\usepackage{soul}

\usepackage[commentmarkup=uwave]{changes}

\usepackage[sectionbib]{bibunits}
\defaultbibliographystyle{apsrev4-1} 
\defaultbibliography{OurRefs} 

\allowdisplaybreaks

\newcommand{\CIT}{\affiliation{Department of Physics, California Institute of Technology, Pasadena, California 91125, USA}}
\newcommand{\CITLab}{\affiliation{LIGO Laboratory, California Institute of Technology, Pasadena, California 91125, USA}}
\newcommand{\CCA}{\affiliation{Center for Computational Astrophysics, Flatiron Institute, New York, NY 10010, USA}}
\newcommand{\AEI}{\affiliation{Max Planck Institute for Gravitational Physics
(Albert Einstein Institute), D-14476 Potsdam, Germany}}
\newcommand{\UMassD}{\affiliation{Department of Mathematics,
    Center for Scientific Computing and Data Science Research,
    University of Massachusetts, Dartmouth, MA 02747, USA}}  
\newcommand{\Monash}{\affiliation{School of Physics and Astronomy
Monash University,
Clayton, VIC 3800, Australia}}  
\newcommand{\OzGrav}{\affiliation{OzGrav: The ARC Centre of Excellence for Gravitational Wave Discovery, Clayton VIC 3800, Australia}}

\definecolor{kcmagenta}{rgb}{0.54, 0.17, 0.88}
\definecolor{shyellow}{rgb}{0.15625, 0.609375, 0.316406}
\definecolor{chorange}{rgb}{0.851, 0.372, 0.007}
\definecolor{tlteal}{rgb}{0,.55,.55}
\definecolor{jcpink}{rgb}{1.0, 0.0, 0.5}
\definecolor{mmgreen}{rgb}{0.0, 0.8, 0.6}
\definecolor{bbsalmon}{rgb}{1.0, 0.47, 0.42}
\definecolor{smgreen}{rgb}{0.26, 0.625, 0.277}

\definechangesauthor[name=Max Isi, color=olive]{MI}

\newcommand{\chieff}{\chi_{\textrm{eff}}}
\newcommand{\chip}{\chi_\mathrm{p}}

\newcommand{\Msun}{\mathrm{M}_\odot} 

\newcommand{\chiperp}{\vec \chi_\perp}
\newcommand{\rhop}{\rho_\mathrm{p}}

\newcommand{\Mref}{M_\mathrm{ref}}

\begin{document}

\title{GW190521: Tracing imprints of spin-precession on the most massive black hole binary}

\author{Simona J.~Miller} \CIT \CITLab 
\author{Maximiliano Isi} \CCA
\author{Katerina Chatziioannou} \CIT \CITLab 
\author{Vijay Varma} \UMassD \AEI
\author{Ilya Mandel} \Monash \OzGrav
\date{\today}

\begin{abstract}

GW190521 is a remarkable gravitational-wave signal on multiple fronts: its source is the most massive black hole binary identified to date and could have spins misaligned with its orbit, leading to spin-induced precession; an astrophysically consequential property linked to the binary's origin.
However, due to its large mass, GW190521 was only observed during its final 3-4 cycles, making precession constraints puzzling and giving rise to alternative interpretations, such as eccentricity.
Motivated by these complications, we trace the observational imprints of precession on GW190521 by dissecting the data with a novel time domain technique, allowing us to explore the morphology and interplay of the few observed cycles.
We find that precession inference hinges on a quiet portion of the pre-merger data that is suppressed relative to the merger-ringdown. 
Neither pre-merger nor post-merger data alone are the sole driver of inference, but rather their combination; in the quasi-circular scenario, precession emerges as a mechanism to accommodate the lack of a stronger pre-merger signal in light of the observed post-merger.
In terms of source dynamics, the pre-merger suppression arises from a tilting of the binary with respect to the observer.
Establishing such a consistent picture between the source dynamics and the observed data is crucial for characterizing the growing number of massive binary observations and bolstering the robustness of ensuing astrophysical claims.
\end{abstract}

\maketitle


\section{Introduction}

At a total mass of ${\sim}150\,\Msun$, GW190521~\cite{GW190521_detection, GW190521_astro} is the current record-holder among massive black hole binaries confidently detected through gravitational waves by LIGO~\cite{aLIGO} and Virgo~\cite{aVirgo}. 
Such high-mass systems are essential probes of the role of hierarchical mergers~\cite{Gerosa:2017kvu,Gerosa:2021mno,Kimball:2020opk,Doctor:2019ruh,Fragione:2021nhb} and pair-instability physics~\cite{Woosley:2016hmi, Woosley:2021xba, Belczynski:2016jno, vanSon:2020zbk, Stevenson:2019rcw, Renzo:2020smh} in binary formation and evolution.
Observationally, massive binaries merge toward the low edge of the detectors' bandwidth and are only detectable for a short time.
One third of the binaries in the the latest gravitational-wave catalog have a median detector-frame total mass $>100\,\Msun$~\cite{GWTC3}, corresponding to ${\sim}5$ signal cycles at more than half a standard deviation above the noise. 
The short duration makes characterizing these signals and inferring astrophysical properties such as spin challenging.

Spin is a key signature of the physics behind angular momentum transport in stellar interiors, black-hole formation, black-hole retention in dense environments, and more~\cite{Mandel:2018hfr,Rodriguez:2016vmx, Rodriguez:2019huv,Gerosa:2018wbw,Farr:2017uvj,Fuller:2019sxi}.
Gravitational-waves provide one of the few ways to measure spins for stellar-mass black holes directly.
Spin components \textit{parallel} to the binary's orbital angular momentum affect the signal duration~\cite{Campanelli:2006uy} and are approximately conserved during the inspiral as the ``effective spin"~\cite{Racine:2008qv,Ajith:2009bn}.
Spin components \textit{perpendicular} to the orbital angular momentum, i.e., \textit{in the orbital plane}, cause the binary to precess, leading to signal modulations as the emission pattern varies relative to the line of sight~\cite{Apostolatos1994,Kidder1993}.
Although typically weak~\cite{Vitale:2014mka, Vitale:2016avz, Shaik:2019dym}, this effect is highly sought-after: spin-induced precession and the associated in-plane spins could differentiate between dynamical and field binary formation, e.g.,~\cite{Mandel:2018hfr, Zevin:2020gbd,Zhang:2023fpp}.

The elusiveness of precession is exacerbated for heavy systems.
The precession timescale can be longer than the observed inspiral for large masses~\cite{Gerosa:2015tea}, making modulations difficult to identify.
The exact imprint of precession on the ensuing merger and ringdown remains poorly understood and analytically intractable, although numerical-relativity and data-analysis studies suggest that imprints do exist~\cite{OShaughnessy:2012iol,Biscoveanu:2021nvg,Hughes:2019zmt,Finch:2021iip,Kamaretsos:2012bs,Hamilton:2023znn};
for example, Ref.~\cite{Biscoveanu:2021nvg} suggests, based on simulations, that high-frequency data typically associated with the merger-ringdown can constrain precession.
This uncertainty makes it difficult to distinguish precession from eccentricity, another highly-valuable binary property~\cite{Romero-Shaw:2020thy, Romero-Shaw:2022fbf, CalderonBustillo:2020xms}.
Interpretation is further complicated by the high sensitivity of the measurability of precession to the system's true parameters~\cite{Biscoveanu:2021nvg, Fairhurst:2019vut,Green:2020ptm,OShaughnessy:2012iol} and the priors~\cite{Olsen:2021qin}.

In light of this, the massive system that sourced GW190521  stands out for its informative precession constraint, when analyzed assuming a quasicircular orbit.
Precession can be quantified by the effective precessing spin $\chip$~\cite{Schmidt:2010it, Schmidt:2012rh, Schmidt:2014iyl} that is motivated by inspiral dynamics.
A value of zero (one) indicates no (maximal) precession.
Under the assumption of a quasicircular orbit, GW190521 has $\chip = 0.68^{+0.25}_{-0.37}$ at 90\% credibility~\cite{GW190521_detection, GW190521_astro}, the largest inferred $\chip$ and the one whose posterior is the most informative to date~\cite{GWTC2,GWTC2.1,GWTC3};\footnote{GW200129, another massive binary~\cite{GWTC3}, has a potentially comparable constraint. When averaged over waveform models, its $\chip$ is unconstrained~\cite{GWTC3}; when restricting to the \texttt{NRSur7dq4}~\cite{Varma:2019csw} model, GW200129 is inferred to be precessing~\cite{Hannam:2021pit}. This interpretation is however complicated by data quality issues~\cite{Payne:2022spz}.}
similar conclusions are reached under alternative parametrizations for precession~\cite{Gerosa:2020aiw,Gangardt:2022ltd}.\footnote{These and further ways to quantify precession are elaborated upon in Appendix \ref{app:precession}.}
The combined high mass and large in-plane spin make GW190521 an essential probe of hierarchical black hole mergers~\cite{Miller:2001ez,Kimball:2020opk,Kimball:2020qyd}, dense stellar environments such as nuclear star clusters~\cite{Rodriguez:2019huv,Gerosa:2019zmo,Fragione:2020han}, active galactic nuclei disks~\cite{McKernan:2014oxa,Mckernan:2017ssq,McKernan:2019beu}, and more~\cite{GW190521_astro, Belczynski:2020bca, Renzo:2020smh,Gondan:2020svr,DeLuca:2020sae,Fishbach:2020qag,Nitz:2020mga}.

The high mass of GW190521 and its few observable cycles open the door to competing astrophysical interpretations.
\citet{Romero-Shaw:2020thy} and \citet{Gayathri:2020coq} find that the data are consistent with eccentricity, though this interpretation is not supported by \citet{Iglesias:2022xfc} and \citet{Ramos-Buades:2023yhy}.
\citet{Gamba:2021gap} propose a hyperbolic capture scenario.
\citet{Nitz:2020mga} suggest a highly asymmetric, but still precessing, binary interpretation.
More exotic explanations include boson stars~\cite{CalderonBustillo:2020fyi} and cosmic strings~\cite{Aurrekoetxea:2020tuw}. 
Any of these alternatives would have important implications if confirmed~\cite{Zevin:2021rtf, Romero-Shaw:2022xko}. 
Additionally, random detector noise can have an outsized impact on the inference of poorly-constrained effects, although \citet{Biscoveanu:2021nvg} and \citet{Xu:2022zza} show that the inference of $\chip$ away from zero in GW190521-like systems cannot be due to Gaussian noise alone.
The fact that full-scale parameter estimation allows for competing interpretations suggests that different physical effects can result in similar observational imprints over GW190521's few cycles. 
Similarly to precession and eccentricity, these imprints are often not analytically tractable.

Toward bolstering the interpretation of massive binaries, it is essential to gain intuition about the observable imprint of physical effects of interest and how their measurability is affected by mismodeling.
Lacking analytical equations for precession in the merger phase, we introduce a novel approach that traces its imprint along the signal and identifies the role of each \textit{cycle} on the $\chip$ constraint.
We dissect the data in the \textit{time-domain} and compare inference between different data subsets. 
We provide a cycle-by-cycle physical picture of source dynamics and explore the interplay of different data regions. 
Our work focuses on the data aspect that drives the inference of precession within a quasi-circular merger scenario. 
Extensions to further physical effects of interest such as orbital eccentricity can be tackled under a similar framework; we leave those to future work.

\section{Methods}

Gravitational-wave parameter estimation is typically conducted in the frequency domain for computational efficiency. 
Leveraging the stationarity of detector noise, noise components at different frequencies are independent which leads to a diagonal covariance matrix in likelihood calculations~\cite{LIGOScientific:2019hgc, Unser:1984}.
However, frequency-domain methods are non-local in time; thus isolating temporal features of source dynamics and their imprint on the data requires nontrivial likelihood modifications~\cite{Zackay:2019kkv,Capano:2021etf}.\footnote{Frequency truncation enables consistency checks~\cite{Cabero:2017avf,Ghosh:2016qgn}, investigations of data-quality issues~\cite{Davis:2022ird, Payne:2022spz}, or alternative studies of the measurability of precession in simulated data~\cite{Biscoveanu:2021nvg}, but this is not equivalent to cuts in time.}
We instead adopt direct time-domain inference to isolate different signal cycles, an approach originally conceived for black hole ringdowns~\cite{Isi:2021iql,Isi:2019aib,Isi:2020tac,Carullo:2019flw}.
We truncate data from LIGO Livingston, LIGO Hanford, and Virgo at different times ranging from $t=-50\,\Mref$ to $50\,\Mref$ with respect to coalescence.\footnote{We define $t$ with respect to geocenter GPS time $1242442967.405764\,\mathrm{s}$. Under geometric units we adopt the median detector-frame remnant mass scale $\Mref = 1.27$\,ms~\cite{GW190521_astro}; in standard units $\Mref = 258.3\,\Msun$. The choice of remnant rather than total mass, was inspired by ringdown analyses~\cite{Siegel:2023lxl}.}
We independently infer the signal properties solely from data before and after each cutoff as well as the full span of data.

We model the signal with the numerical relativity surrogate model \texttt{NRSur7dq4}~\cite{Varma:2019csw}, which assumes quasi-circular orbits and includes precession and higher-order modes. 
Within its region of validity, \texttt{NRSur7dq4} displays the lowest mismatches against numerical relativity among existing models~\cite{Varma:2019csw}. 
We adapt the time-domain inference code from~\citet{Isi:2020tac} and sample the multidimensional posterior for the binary masses, spin magnitudes and tilt angles, azimuthal inter-spin angle, azimuthal precession cone angle, inclination, luminosity distance, and phase of coalescence.  
The time of coalescence, right ascension, declination, and polarization angle are fixed for computational efficiency.\footnote{We have verified that these choices do not affect our conclusions. All parameter estimation settings, priors, and consistency checks are given in Appendix \ref{app:PEsettings}}.

We report precession constraints using the canonical effective precessing spin, $\chip$~\cite{Schmidt:2010it, Schmidt:2012rh, Schmidt:2014iyl}:
\begin{equation}\label{eqn:chip}
    \chi_\mathrm{p} = \mathrm{max}\left[\chi_1 \sin\theta_1,
    \left(\frac{3+4q}{4+3q}\right) q \,\chi_2 \sin\theta_2\right]\in [0,1)  \,.
\end{equation}
Here $\chi_i \in [0,1)$ are the dimensionless spin magnitudes and $\theta_i$ are the tilt angles between the spin and orbital angular momentum vectors. 
Subscripts $i \in \{1,2\}$ denote each black hole with mass $m_i$ and $q \equiv m_2/m_1 \leq 1$.

\section{Results}

\begin{figure*}
    \includegraphics[width=0.95\textwidth]{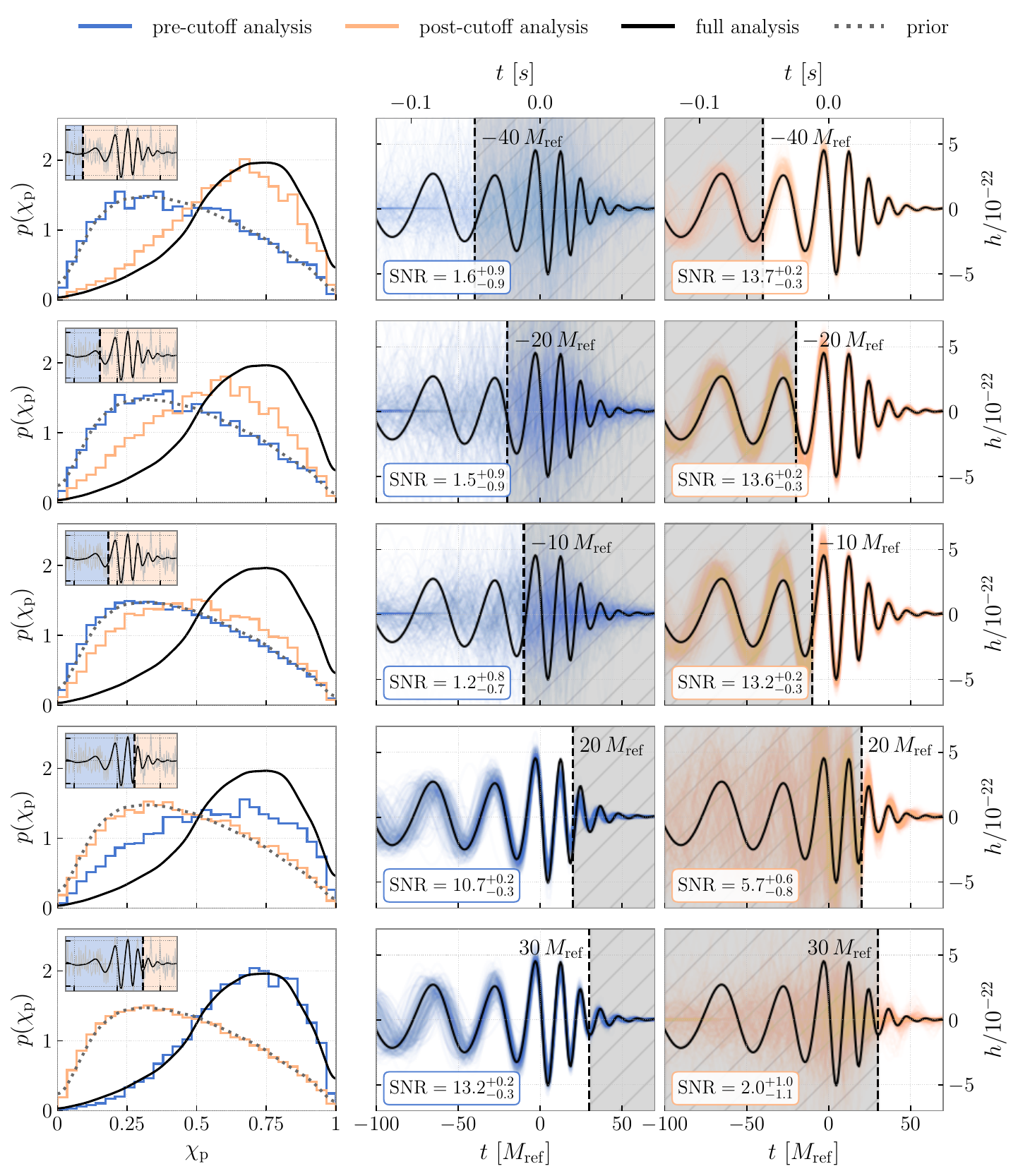}
    \caption{
    Evolution of GW190521 inference for representative cutoff times $t \in \{-40, -20, -10, 20, 30\}\,\Mref$ from the time of coalescence (top to bottom; vertical black dashed lines where applicable). 
    \textit{Left:} Posterior for $\chip$ from the pre- (blue), post-cutoff (orange), and full (black solid) analysis and prior (gray dotted). 
    The inset shows the whitened maximum-posterior waveform (with $\chip =0.62$, detector-frame total mass $M = 267\,M_\odot$, and $q=0.89$) from the full analysis (black) along with the whitened LIGO Livingston data (gray).
    The blue/orange shaded regions highlight the data informing the same-color $\chip$ posterior.
    \textit{Center and Right:} Waveform reconstruction draws for LIGO Livingston from the pre- (center, blue) and post-cutoff (right, orange) analyses and maximum-posterior waveform  for the full analysis (black). 
    Median and 90\% credible intervals for the matched-filter network SNRs are given in-figure.
    Gray shading denotes data excluded from each analysis.
    See Ref.~\cite{gw190521_animation1} for an animation of this figure including more cutoff times. 
    }
    \label{fig:figure_01}
\end{figure*}

Figure~\ref{fig:figure_01} shows the inferred GW190521 properties from data before (blue) and after (orange) five representative cutoff times (vertical lines) as well as the full signal (black) for comparison.\footnote{Results for further cutoff times are included in our accompanying Github repository~\cite{github_release}.}
Insets in the left panels visualize the truncation in LIGO Livingston, selected as the detector in which GW190521 is the loudest. 
The left column shows the posterior for $\chip$ at a reference frequency of 11\,Hz~\cite{GW190521_astro,GW190521_detection}.
For the earliest cutoff time (top row), the post-$t = -40\,\Mref$ $\chip$ posterior is almost identical to that of the full analysis, while the pre-cutoff one is identical to the prior. 
This is due to the fact that the post-cutoff analysis includes the full available signal, while none of it is contained in the pre-cutoff data (see inset).
As the cutoff moves to later times (top to bottom), the pre- and post-cutoff posteriors gradually exchange places as the data preceding each cutoff become more informative and the data following become less so. 

\textit{The ${\sim}40\,$ms between $t=-40\,\Mref$ and $-10\,\Mref$ are crucial to constrain precession for GW190521}.
This region roughly corresponds to the final cycle before the onset of merger. 
The $\chip$ posterior obtained from data after $t=-40\,\Mref$ (first row, orange) is consistent with that from the full analysis, i.e., precession \textit{is} constrained.\footnote{By ``constrained," we specifically mean ``visibly different from the prior."}
On the other hand, data after $t=-10\,\Mref$ (third row, orange), result in a $\chip$ posterior that is nearly identical to the prior, i.e., uninformative.
Between these times, the posterior shifts smoothly between the full measurement and the prior; e.g., the post-$t=-20\,\Mref$ analysis (second row, orange). 
The reduction in the signal-to-noise ratio (SNR) from excluding data is negligible between $-40\,\Mref$ and $-10\,\Mref$ suggesting this qualitative change in precession inference is not due to an SNR drop, see Fig.~\ref{fig:figure_08} in Appendix \ref{app:snr}.  

\textit{Neither the inspiral nor the merger/ringdown data alone are fully responsible for precession constraints in GW190521.} 
The data both pre- and post-$t=-10\,\Mref$ alone are uninformative about precession (third row, orange and blue).
Moreover, the pre-$t=30\,\Mref$ analysis that excludes the final ringdown cycle (fifth row, blue) is consistent with the full analysis. 
It is therefore not solely the final pre-merger cycle that informs precession, but rather its 
combination with the subsequent $2$ merger and early-ringdown cycles.
This does not rule out ringdown imprints of precession that are too weak to discern at this SNR or with this waveform.

\begin{figure}
    \centering
    \includegraphics[width=\linewidth]{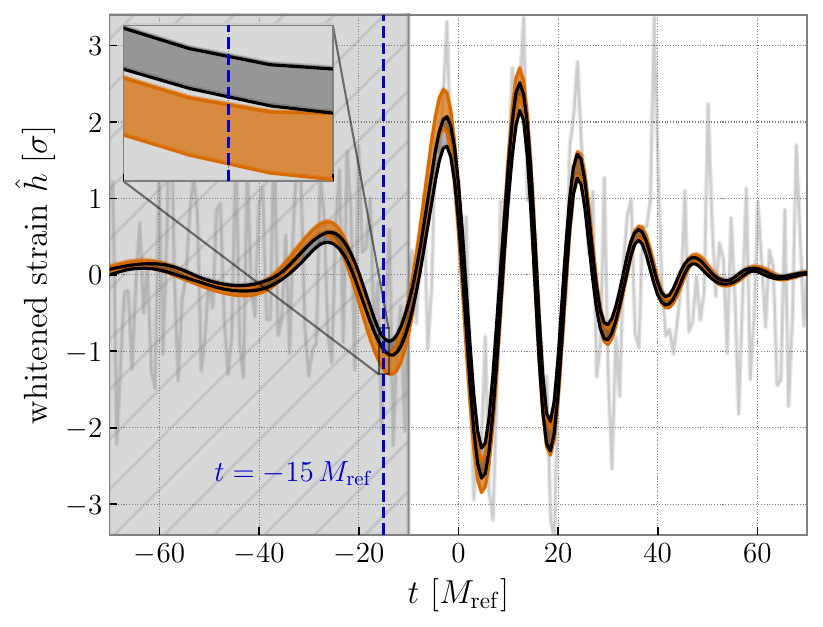}
    \includegraphics[width=\linewidth]{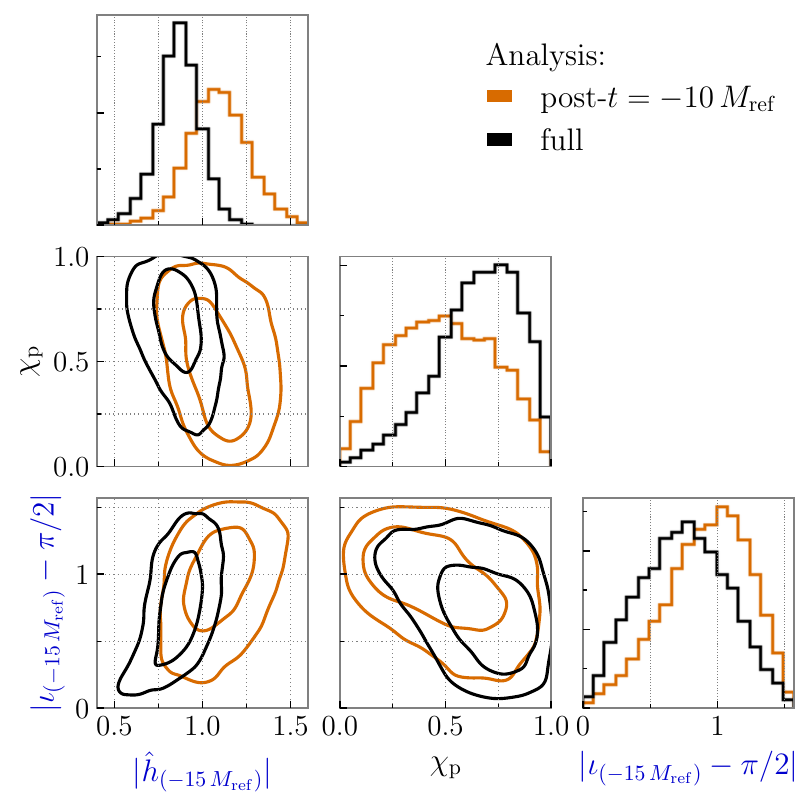}
    \caption{Results from the full (black) and from the post-$t=-10\,\Mref$ analysis (orange) that are informative and uninformative about precession respectively.
    \textit{Top:} 50\% credible intervals for the whitened waveforms in LIGO Livingston in units of standard deviations of the noise.
    Data are plotted in gray.  
    Gray shading denotes data excluded from the post-$t = -10\,\Mref$ analysis.
    The inset zooms in around
    $t=-15\,\Mref$ (blue dashed line), the minimum of the final pre-merger cycle.
    \textit{Bottom:} Posteriors for $\chip$, the absolute value of the whitened strain $|\hat h|$, and the inclination angle relative to edge-on configurations $|\iota-\pi/2|$. 
    Quantities labeled in blue are plotted at $t=-15\,\Mref$.
    Contours denote 50\% and 90\% credible regions. 
    The whitened strain is anticorrelated with $\chip$ and correlated with $|\iota - \pi/2|$. 
    Large $\chip$ is paired with smaller pre-merger signal and more edge-on configurations.
    }
    \label{fig:figure_02}
\end{figure}

The center and right columns of Fig.~\ref{fig:figure_01} investigate \textit{features} of the waveforms. The blue and orange waveforms are informed only by data in the unshaded regions and extended coherently into the shaded regions.     
As progressively less data are analyzed (center bottom to top, right top to bottom), the waveform reconstructions agree less with the full-analysis waveform, eventually becoming incoherent.
The right column reveals the morphological imprint of precession on the signal during the transition from an informative $\chip$ posterior (first row) to the prior (third row). 
When the $\chip$ inference returns (close to) the prior, the final pre-merger cycle is extrapolated to be \textit{larger} than when $\chip$ is constrained to take higher values, c.f., the waveform peak at $t\sim-30\,\Mref$ and trough at $t\sim-15\,\Mref$. 
Again there is a progression: the post-$t = -40\,\Mref$ inferred waveforms (orange) are consistent with the full analysis (black), while the final pre-merger cycle subtly increases in strength toward post-$t = -10\,\Mref$ (top row to third row).

\begin{figure}
    \centering
    \includegraphics[width=\linewidth]{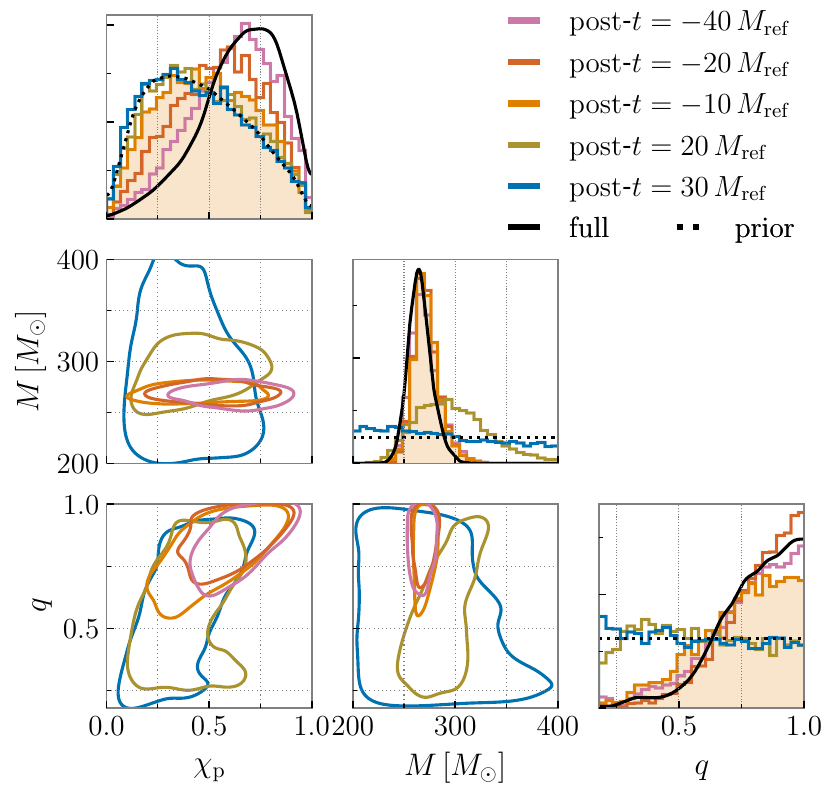}
    \caption{Posteriors for $\chip$, $M$ (detector frame total mass), and $q$ (mass ratio) for the same cutoff times as Fig.~\ref{fig:figure_01}.
    Posteriors from the full analysis and priors are plotted in black solid and dotted respectively. 
    Contours denote 50\% credible regions.
    Information about spin-precession is lost post-$t = -10\,\Mref$ (shaded light orange), while the total mass and mass ratio posteriors are informative at post-$t = 20\,\Mref$ (green).
    See Ref.~\cite{gw190521_animation2} for an animation showing corner plots for mass and spin parameters for the pre- and post-cutoff analyses at more cutoff times.
    }
    \label{fig:figure_03}
\end{figure}

To further explore the pre-merger waveform suppression, we compare the full analysis (black) in which precession \textit{is} constrained and the post-$t=-10\,\Mref$ analysis where the data are uninformative about precession in Fig.~\ref{fig:figure_02}.
In order to focus on waveform features that are informative compared to the noise, we plot the whitened waveform.\footnote{Whitened waveforms are obtained by dividing the Fourier-domain waveform by the noise amplitude spectral density and then inverse Fourier-transforming. See Appendix \ref{app:whitening} for details.}
In the top panel the grayed-out region denotes data available to the full analysis but not the post-$t=-10\,\Mref$ one. 
The inset focuses around $t=-15\,\Mref$, an extremum of the final pre-merger cycle. 
The reconstructions are inconsistent at the 50\% credible level, with the post-$t=-10\,\Mref$ analysis resulting in a larger amplitude (in absolute value).
This inconsistency only occurs at the extrema of the final pre-merger cycle, i.e.~the peak around $t\sim-30\,\Mref$ and trough at $t\sim-15\,\Mref$, see Fig.~\ref{fig:figure_09} in Appendix \ref{app:snr}.

The bottom panel shows marginal posteriors for select quantities: the effective precessing spin $\chip$, the absolute value of the whitened strain $|\hat h|$ (units of standard deviations $\sigma$ of the noise), and the difference between $\iota$---the angle between the direction of maximum signal emission and the line of sight~\cite{Boyle:2011gg}---and $\pi/2$ at $t=-15\,\Mref$.
As expected from Fig.~\ref{fig:figure_01}, $\chi_p$ and $|\hat h |$ are anticorrelated (albeit weakly):
when the full data are analyzed (black), the final cycle is constrained to be weaker, resulting in a larger $\chip$;
when this cycle is excluded from the analysis (orange), the extrapolated waveform is not required to have such a small value, obviating the need for a higher $\chip$.
In summary, \textit{precession in GW190521 is informed by a suppression of the gravitational-wave in the observed waveform's last cycle before merger}.
This is also the region in which the waveform is overall quietest: the whitened signal is less than $1\,\sigma$ above the noise, compared to the subsequent merger cycles that are over $2\,\sigma$. 
 
The origin of the signal suppression can be attributed to the evolution of the emission direction.
As a binary precesses, the angle between the dominant emission direction and the line-of-sight evolves, changing the amplitude of the observed signal.
Systems with $\iota \sim \pi/2$ are quieter than those with $\iota \sim 0$ or $\sim \pi$.
To explore these dynamics, we plot the absolute difference between the inclination angle and $\pi/2$ at the time $t = -15\,\Mref$ in Fig.~\ref{fig:figure_02}.
As expected, $\chip$ and $|\iota - \pi/2|$ are anti-correlated, while $|\hat h|$ and $|\iota - \pi/2|$ are correlated.
When precession is constrained, $\iota$ is found to be closer to edge-on at the last pre-merger cycle than when precession is unconstrained, leading to a suppressed cycle and smaller $|\hat h|$.

We investigate the impact of data truncation on other source parameters in Fig.~\ref{fig:figure_03}.
Information about $\chip$, $M$ and $q$ is not lost or gained in lockstep as a function of cutoff time. 
At post-$t = -20\,\Mref$ (dark orange), the $\chip$ posterior shifts away from the full analysis posterior;
however, for $M$ and $q$, this does not happen until multiple cycles later. 
Post-$t = -10\,\Mref$ (shaded light orange), i.e., at the end of the final pre-merger cycle, the $\chip$ posterior is close to the prior, while the detector-frame total mass $M$ and mass ratio $q$ both resemble the posteriors from the full analysis.
Thus, the lack of an informative $\chip$ posterior post-$t = -10\,\Mref$ does not simply arise from poor parameter constraints over-all due to lower SNR; rather, the suppression of the final pre-merger cycle is informative \textit{specifically} about precession(under the assumption that the system has a quasi-circular orbit).
We also confirm that the $\chip$ inference is not driven by a conditional measurement based on the typically better-measured aligned spins in Appendix \ref{app:precession}.

\section{Conclusions}

Precession inference for the massive, distant binary black hole signal GW190521 is subtle. 
It originates from contrasting a ${\sim}40~\mathrm{ms}$ slice of data from the final pre-merger cycle between $t = -40\,\Mref$ and $-10\,\Mref$ with the loud merger cycles following it.
The merger of GW190521 is $2$ loud cycles that reach $2.5\,\sigma$ above the noise and is informative about the source's masses.
However, precession is only constrained away from the prior when the merger is observed in tandem with the final pre-merger cycle, which does not rise more than $1\,\sigma$ above the noise.
The measurement is linked to a relative suppression of the aforementioned final cycle, caused by the binary tilting toward an edge-on configuration due to precession. 
This picture qualitatively agrees with the interpretation posited in Ref.~\cite{GW190521_detection} by comparing precessing and spin-aligned waveform reconstructions for the full signal, and is supported by simulations~\cite{CalderonBustillo:2020xms}.

\citet{Siegel:2023lxl} carried out a complementary study seeking a description of the the GW190521 ringdown consistent with the \texttt{NRSur7dq4} full analysis.%
\footnote{\citet{Capano:2021etf,Capano:2022zqm} provided an alternative interpretation based on waveform models of the \textsc{Phenom} family~\cite{Pratten:2020ceb,Estelles:2021jnz}.}
Ringdown mode content encodes information about the preceeding binary dynamics~\cite{Kamaretsos:2012bs, Borhanian:2019kxt,Abedi:2023kot}, meaning it is (in theory) possible to identify signatures of precession in the ringdown.
\citet{Siegel:2023lxl} found support for the presence of at least two modes; consistency with \texttt{NRSur7dq4} suggests a configuration including the 220 and 210 fundamental modes.
A large 210 mode amplitude could be expected under strong precession~\cite{Siegel:2023lxl}.
The fact that past GW190521 ringdown-only analyses cannot unequivocally infer precession is consistent with our finding that a post-peak analysis is not sufficient to constrain precession.

Our study highlights the delicate nature of precessional imprints on observed signals, providing a new view of spins in massive systems beyond the frequency domain~\cite{Biscoveanu:2021nvg,Xu:2022zza,GW190521_detection}. For GW190521, the difference between the most informative precession inference to date and the prior boils down to a single, quiet pre-merger cycle that needs to be measured to better than half a standard deviation, \textit{cf}., the difference between the black and orange waveforms in Fig.~\ref{fig:figure_02}.
This is quantitatively in agreement with the conclusions of \citet{Payne:2022spz} who explored the impact of data quality on our ability to obtain an unbiased measurement at that level. 

Our novel time-domain approach of tracing the observational imprint of interesting physical effects cycle-by-cycle can provide physical intuition about how key source properties are inferred in relation to observed data features.
In anticipation of further massive observed signals, such a correspondence between source dynamics and observed data can help pinpoint the most informative data in order to assess data quality and waveform systematics, and enable us to morphologically study competing physical interpretations that are likely to keep arising.


\section*{Data Availability}

Posterior samples from all our pre- and post-cutoff analyses -- including those from additional cutoff times not included in the main text -- are available on Zenodo at Ref.~\cite{zenodo_release}. 
We release posteriors for all cutoff times in intervals of $10\,\Mref$, from $t=-50\,\Mref$ to $50\,\Mref$.  
Cutoff times in intervals of $2.5\,\Mref$ between $t=-30\,\Mref$ and $20\,\Mref$ were additionally explored to more finely resolve the transition between informative and uninformative $\chip$ posteriors. 

Scripts to generate the waveform reconstructions and inclination angles are on Github at Ref.~\cite{github_release}, as are notebooks to plot all the figures appearing in the text.
The repository additionally contains animations showing results from for all pre- and post-cutoff analyses in intervals of $t = 2.5\,\Mref$:

\begin{itemize}

    \item Ref.~\cite{gw190521_animation1}: Animation of $\chip$ posteriors, whitened reconstructions, and colored reconstructions for all time slices; similar to Fig.~\ref{fig:figure_01}.
    
    \item Ref.~\cite{gw190521_animation2}: Animation of a corner plot for $\chip$, $\chieff$, $M$, and $q$ at all time slices; similar to Figs.~\ref{fig:figure_03} and \ref{fig:figure_06}.

\end{itemize}


\acknowledgements

We thank Harrison Siegel for helpful discussions on ringdown analyses of GW190521, Sylvia Biscoveanu for insights about inference on high-mass gravitational-wave sources, Sophie Hourihane for assistance whitening gravitational-wave signals, Davide Gerosa for insight on alternative measures of precession, and Will Farr for insights about time-domain inference.
We also extend thanks to Jacob Lange, Christopher Berry, Carlos Lousto, Juan Calderon Bustillo, and Salvatore Vitale for their helpful comments on our manuscript.
 
This material is based upon work supported by NSF's LIGO Laboratory which is a major facility fully funded by the National Science Foundation.
S.M.~and K.C.~were supported by NSF Grant PHY-2110111. 
K.C. acknowledges support from the Sloan Foundation.
The Flatiron Institute is funded by the Simons Foundation.
V.V.~acknowledges support from NSF Grant No. PHY-2309301, and the European
Union’s Horizon 2020 research and innovation program under the Marie
Skłodowska-Curie grant agreement No.~896869.
I.M.~is a recipient of the Australian Research Council Future Fellowships (FT190100574).
The authors are grateful for computational resources provided by Cardiff University and supported by STFC grant ST/I006285/1.

Software: \texttt{emcee}~\cite{emcee}, \texttt{LALSuite}~\cite{lalsuite}, \texttt{numpy}~\cite{numpy}, \texttt{scipy}~\cite{scipy},
\texttt{h5py}~\cite{h5py}, \texttt{matplotlib}~\cite{matplotlib}, \texttt{seaborn}~\cite{seaborn}, \texttt{ringdown}~\cite{Isi:2019aib, Isi:2021iql}, \texttt{gwtools}~\cite{gwtools}.

\appendix


\section{Parameter estimation settings and priors}
\label{app:PEsettings}

\begin{figure*}
    \centering
    \includegraphics[width=\linewidth]{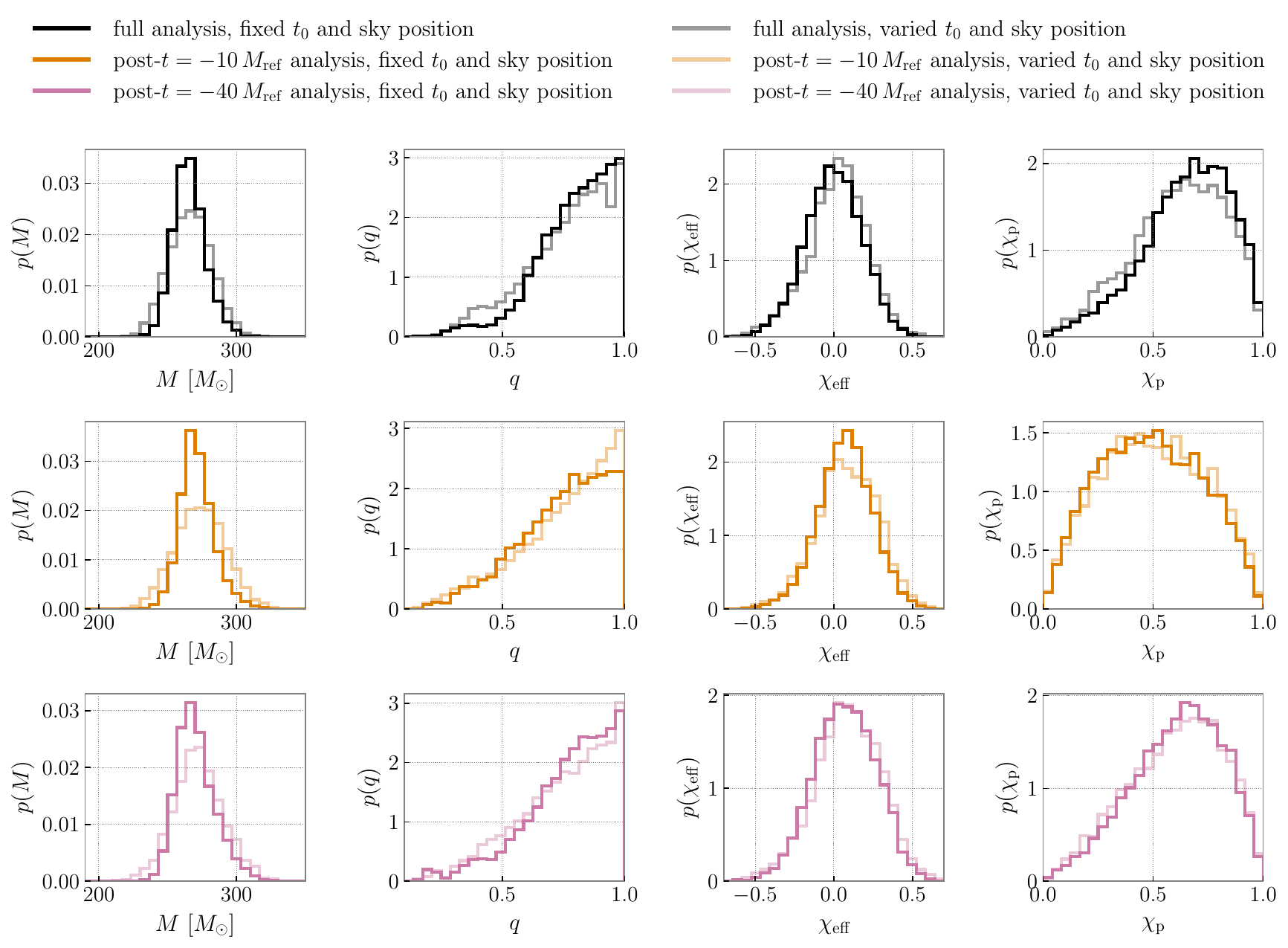}
    \caption{
    Posteriors for the detector-frame total mass $M$, mass ratio $q$, effective spin $\chieff$, and effective precessing spin $\chip$ for parameter estimation with (lighter color) and without (darker color) sampling over time of coalescence $t_0$ and sky position -- referring to the right ascension, declination, and polarization angle -- for the full signal (black; top row), the post-$t=-10\,\Mref$ (orange; middle row), and the post-$t=-40\,\Mref$ (pink; bottom row). 
    All results are consistent, the only difference being a slightly wider total mass posterior.}
    \label{fig:figure_04}
\end{figure*}

\begin{figure*}
    \centering
    \includegraphics[width=\linewidth]{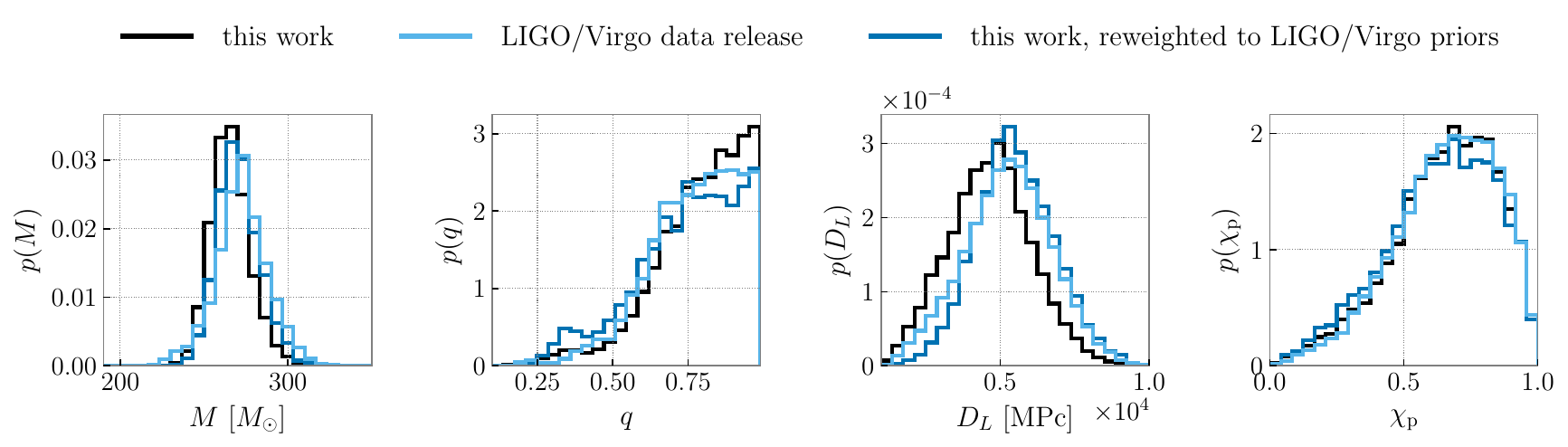}
    \caption{Posteriors for the detector-frame total mass $M$, mass ratio $q$, luminosity distance $D_L$, and effective precessing spin $\chip$ from this work (black), Ref.~\cite{GW190521_detection} (LIGO/Virgo data release; light blue), and this work reweighted to the priors of Ref.~\cite{GW190521_detection} (dark blue).
    Our inference is consistent with Ref.~\cite{GW190521_detection} under the same priors. 
    }
    \label{fig:figure_05}
\end{figure*}

During parameter estimation, we sample the masses, spin magnitudes and tilt angles, azimuthal inter-spin angle, azimuthal precession cone angle, inclination angle, luminosity distance, and phase of coalescence. 
The time of coalescence, right ascension, declination, and polarization angle are fixed for computational efficiency; we have ensured that further sampling over these parameters does not alter our conclusions, as is shown in Fig.~\ref{fig:figure_04}.
Priors for all parameters are given in Table~\ref{tab:priors}.
We use slightly different mass and distance priors than Ref.~\cite{GW190521_detection} but -- as we show in Fig.~\ref{fig:figure_05} -- obtain consistent results when analyzing the full signal.\footnote{We use the \texttt{NRSur7dq4}~\cite{Varma:2019csw} posterior samples (specifically \texttt{GW190521\_posterior\_samples.h5}) released by LIGO/Virgo at Ref.~\cite{GW190521_PE} for this comparison.}

Parameter estimation settings are listed in Table~\ref{tab:run_params}.
All analyses make use of the three detector network of LIGO Livingston, LIGO Hanford, and Virgo, using strain and event power spectral densities publicly available on the Gravitational-wave Open Science Center \cite{opendata_O1O2,opendata_O3}.\footnote{Specifically we download the ``32\,sec, 16\,KHz" strain data uploaded to Ref.~\cite{GW190521_gwosc}.}
We use a lower and reference frequency of 11\,Hz, a maximum frequency of 1024\,Hz, and a sampling rate of 2048\,Hz.\footnote{We opt for a minimum frequency of 11\,Hz which is consistent with Refs.~\cite{GWTC2,GWTC3}, but not the more recent Ref.~\cite{GWTC2.1}.}
Each analysis conducted \textit{before} a given cutoff time begins at the GPS time at geocenter $1242442966.9077148$\,s; those conducted \textit{after} a given time all end at $1242442967.607715$\,s.
To translate the cutoff times from geocenter to the times at the three detectors, we use a right ascension of $\alpha=6.075$\,rad, a declination of $\delta=-0.8$\,rad, and polarization angle of $\psi=2.443$\,rad.

\begin{table}
    \centering
    \renewcommand{\arraystretch}{1.5}
    \begin{tabular}{ >{\raggedright}p{0.44\columnwidth} p{0.15\columnwidth} p{0.26\columnwidth} }
        \hline\hline
         Parameter & Symbol & Value\\
         \hline
          Detector-frame total mass & $M(1+z)$ &  $U(200, 500)\,M_\odot$ \\
          Mass ratio & $q$ &  $U(0.17, 1)$ \\
          Primary spin magnitude & $\chi_1$ &  $U(0, 1)$ \\
          Secondary spin magnitude & $\chi_2$ &  $U(0, 1)$ \\
          Primary spin tilt & $\theta_1$ &  isotropic \\
          Secondary spin tilt & $\theta_2$ &  isotropic \\
          Azimuthal inter-spin angle & $\phi_{12}$ & $U(0, 2\pi)$\\
          Azimuthal cone precession angle & $\phi_{JL}$ & $U(0, 2\pi)$\\
          Inclination angle & $\iota$ & isotropic\\
          Luminosity distance & $d_{L}$ & $U(10^3, 10^4)$\,Mpc\\
          Phase of coalescence & $\varphi$ & $U(-\pi,\pi)$\\

         \hline\hline
        \end{tabular}
        \caption{Priors used in parameter estimation. $U(a,b)$ means uniform between $a$ and $b$. An isotropic prior for an angle $x$ means that the prior on $\cos x$ is uniform between $-1$ and $1$. Right ascension, declination, polarization, and time of coalescence are fixed to the values given in Table~\ref{tab:run_params}.}
    \label{tab:priors}
\end{table}
\begin{table}
    \centering
    \renewcommand{\arraystretch}{1.5}
    \begin{tabular}{ >{\raggedright}p{0.38\columnwidth} p{0.1\columnwidth} p{0.37\columnwidth} }
        \hline\hline
         Parameter & Symbol & Value\\
         \hline
         Start GPS time of ``before" segments~~ & $t_\mathrm{start}$ &  1242442966.907715\,s \\
         End GPS time of ``after" segments & $t_\mathrm{end}$ &  1242442967.607715\,s \\
         Coalescence GPS time  & $t_{0}$ & 1242442967.405764\,s \\
         Mass-time scaling relation  & $1\,\Mref$ &  0.00127\,s \\
         Right ascension & $\alpha$ &  6.075\,rad \\
         Declination  & $\delta$ &  -0.800\,rad \\
         Polarization angle & $\psi$ & 2.443\,rad \\
         Minimum frequency  & $f_\mathrm{min}$ &  11\,Hz \\
         Maximum frequency  & $f_\mathrm{max}$ &  1024\,Hz\\
         Sampling rate & $f_\mathrm{samp}$ &  2048\,Hz\\
         Reference frequency  & $f_\mathrm{ref}$ & 11\,Hz \\
         \hline\hline
        \end{tabular}
        \caption{Settings for the time-domain parameter estimation for GW190521. The pre-cutoff analyses all begin at $t_\mathrm{start}$, while the post-cutoff ones end at $t_\mathrm{end}$. These times are calculated with respect to the geocenter time $t_{0}$ and then shifted between detectors using the extrinsic parameters $\alpha$, $\delta$, and $\psi$.}
    \label{tab:run_params}
\end{table}

\section{Whitening Waveforms}
\label{app:whitening}

Whitened waveforms are obtained by dividing the Fourier-domain waveform by the noise amplitude spectral density and then inverse Fourier-transforming.
In the frequency domain, the whitened waveform $\hat h (f)$ is obtained from the original waveform $h(f)$ by:
\begin{equation}
    \hat h (f) = \sqrt{\frac{f_s}{2\,S_n(f)}} ~ h (f)
\end{equation}
where $S_n(f)$ is the power spectral density and $f_s$ is the sampling rate of the data.
Though we sample the data at $2048\,\mathrm{Hz}$ for inference, we use $1024\,\mathrm{Hz}$ when plotting whitened waveforms for direct comparison to Ref.~\cite{GW190521_detection}.


\begin{figure}
    \centering
    \includegraphics[width=\linewidth]{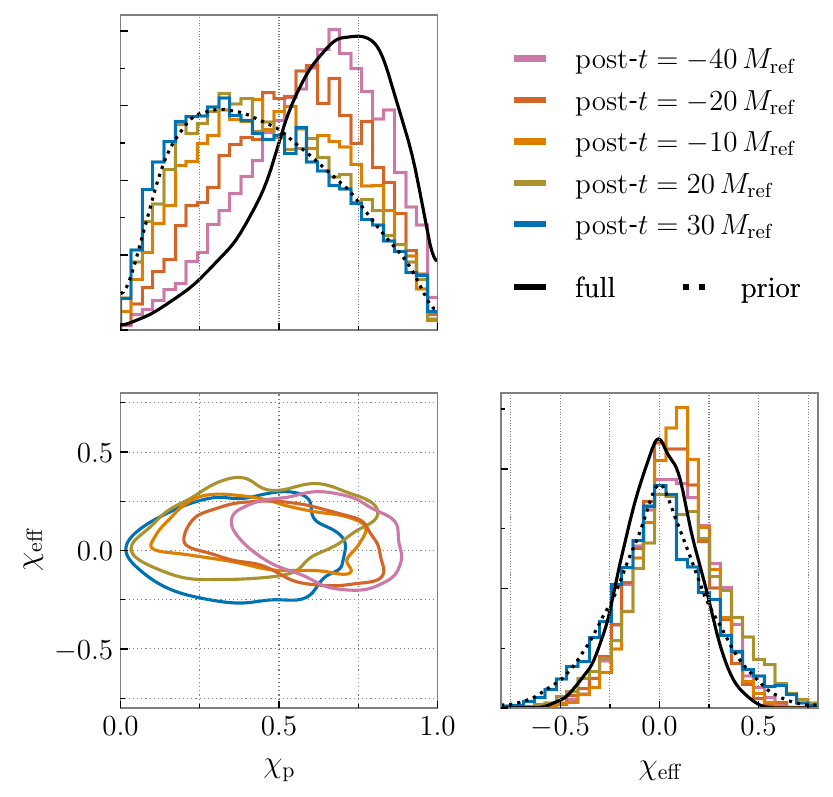}
    \caption{Posteriors for $\chip$ and $\chieff$ for GW190521 informed by data after five representative cutoff times: $t = \{-40, -20, -10, 20, 30\}\,\Mref$ in pink, red, orange, green, blue respectively, compared to the full signal analysis (solid) and the prior (dotted) in black.
    Contours of the 50\% credible regions are shown for the two-dimensional posteriors.
    The two spin parameters are not correlated, meaning that inferences on $\chip$ at the various cutoff times do not simply arise from conditional priors from $\chieff$.
    }
    \label{fig:figure_06}
\end{figure}

\begin{figure}
    \centering
    \includegraphics[width=\linewidth]{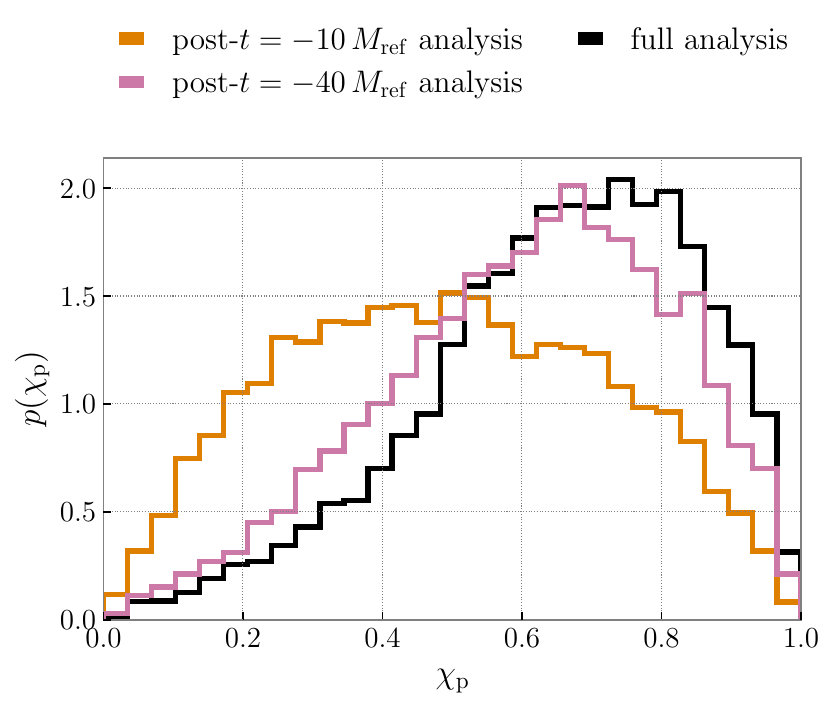}
    \includegraphics[width=\linewidth]{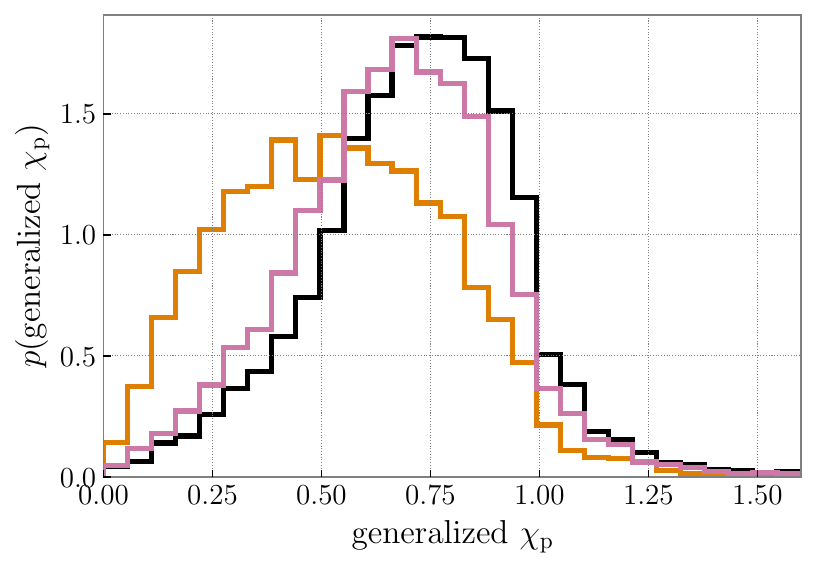}
    \includegraphics[width=\linewidth]{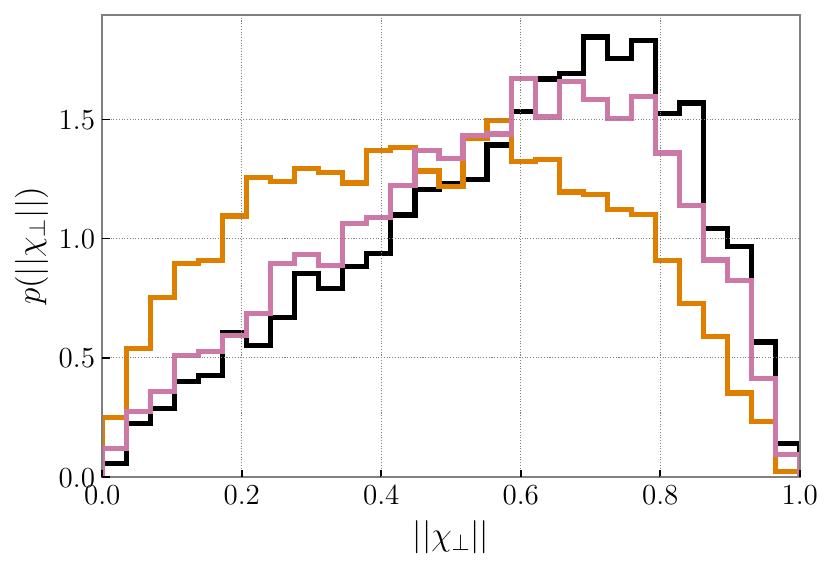}
    \caption{From top to bottom: posteriors for $\chip$, generalized $\chip$, and $||\chiperp||$ for the full analysis (black), and the post-$t = -40\,\Mref$ (pink) and post-$t = -10\,\Mref$ (orange) analyses. All parametrizations of precession lead to qualitatively similar conclusions about spin inference and GW190521.}
    \label{fig:figure_07}
\end{figure}

\section{Quantifying precession}
\label{app:precession}

Spins aligned with the orbit are characterized at leading order by the effective spin $\chieff$, 
\begin{equation}\label{eqn:chieff}
    \chi_\mathrm{eff} = \frac{\chi_1 \cos\theta_1 + q \, \chi_2 \cos\theta_2}{1+q} \in (-1,1)  \,,
\end{equation}
which is conserved to at least the second post-Newtonian order in the inspiral~\cite{Racine:2008qv,Ajith:2009bn}. 
Precession -- stemming from mis-aligned spins -- is typically characterized via the ``effective precessing spin" parameter~\cite{Schmidt:2014iyl}, as defined in Eq.~\eqref{eqn:chip} in the main text.
Posteriors for $\chieff$ versus $\chip$ for five representative cutoff times (compared to the full signal posterior and the prior) are shown in Fig.~\ref{fig:figure_06}.
Though typically $\chieff$ is better constrained than $\chip$, this is not the case for GW190521: $\chieff$ is never inferred to be far from its prior with only slight deviations from the prior that occur at different times from when we best constrain $\chip$.
Furthermore, because the $\chip$ and $\chieff$ are not correlated at any time slice, inferences of $\chip$ -- and thus our main conclusions -- do not arise from a conditional measurement driven by $\chieff$~\cite{Gangardt:2022ltd}.
In other words, we are not just identifying differences in the spins from $\chieff$ that then propagate into differences in $\chip$, but rather we are directly identifying differences in the in-plane spin components' inference. 

However, $\chip$ is a parameter motivated by inspiral precession dynamics~\cite{Schmidt:2010it, Schmidt:2012rh, Schmidt:2014iyl}. 
The angles $\theta_i$ associated with each black hole's spin vector -- used to calculate $\chip$ -- are defined with respect to the binary's orbital angular momentum, a quantity that becomes meaningless during the merger as the two black holes cease orbiting each other.
Indeed, the canonical equations describing precession dynamics (Eq.~(11) of Ref.~\cite{Apostolatos1994}) are only accurate through the second post-Newtonian order, an expansion which is only valid when the black holes are moving substantially slower than the speed of light, i.e.~during the inspiral~\cite{Apostolatos1994,Kidder1993,Blanchet:2013haa,Arun:2008kb,Buonanno:2009zt}.
Once entering the merger, precession becomes no longer analytically tractable.

Alternative parameterizations of precession have been proposed, though they are all still inspired by intuition gained from the inspiral precession equations.
\citet{Gerosa:2020aiw} note that in the definition of $\chip$, some -- but not all -- of the  precession-timescale oscillations are averaged over. 
To rectify this and retain all variation over the precessional timescale~\cite{Gerosa:2015tea}, they define a ``generalized $\chip$" in their Eq.~(15). 
Defined on the range $[0,2)$, the generalized $\chip$ has the advantage that it can help distinguish between binaries in which one versus both spins are precessing. 
While binaries with generalized~$\chip <1$ can have one or both spins precessing, the case in which generalized~$\chip >1$ only arises in the both-precessing scenario.
\citet{Thomas:2020uqj} amend the fact that $\chip$ does not accurately account for higher order multi-polar modes by defining the ``effective precession spin vector" $\chiperp$ --  their Eq.~(9). This parameter accounts for more degrees of freedom and could facilitate a better representation of precession in the strong-field regime. 
The magnitude of the effective precession spin vector $||\chiperp||$ is a scalar measure of precession, analogous to the traditional $\chip$.
A third way to quantify the observability of spin-precession is through the precessional SNR $\rhop$, as defined in Fairhurst \textit{et al.}~\cite{Fairhurst:2019srr, Fairhurst:2019vut}.
Based on the idea that precession is only inferred when two gravitational-wave harmonics are observed, $\rhop$ is related to the SNR in the second most significant waveform harmonic, defined in Eq.~(39) of Ref.~\cite{Fairhurst:2019vut}.

Posteriors for generalized $\chip$ and $||\chiperp||$ are plotted in Fig.~\ref{fig:figure_07} for the post-$t = -40\,\Mref$ (pink) and $-10\,\Mref$ (orange) analyses, compared to the full analysis posteriors (black). 
We do not plot $\rhop$, as the current code to compute it~\cite{pe_summary} necessitates a frequency domain waveform, which \texttt{NRSur7dq4} is not~\cite{Varma:2019csw}; \citet{Hoy:2021rfv} presented $\rhop$ constraints for GW190521 using a different set of waveform models.
Both gen.~$\chip$ and $||\chiperp||$ follow the same trend as the traditional $\chip$, shown in the top panel of Fig.~\ref{fig:figure_07} for comparison. 
The inference of precession is closer to that from the full signal in the post-$t=-40\,\Mref$ analysis than the $t=-10\,\Mref$. 

The above quantities are all defined based on the inspiral dynamics; precession representations motivated by merger dynamics are not available to the best of our knowledge.
Furthermore, all of the above quantities, as well as $\chip$ itself, are quoted at the reference frequency of 11\,Hz.
Since quantifications of precession are time varying--albeit slowly~\cite{Schmidt:2010it}--we expect their measurement to change as a function of reference frequency. 
However, crucially we do not expect the choice of reference frequency to affect our conclusions.
As long as the same reference frequency is used for all pre-/post-cutoff analyses, the posteriors for $\chip$ (and related quantities) are expected to vary with the cutoff time in a similar way to the results presented in Fig.~\ref{fig:figure_07}.

\begin{figure}
    \centering
    \includegraphics[width=\linewidth]{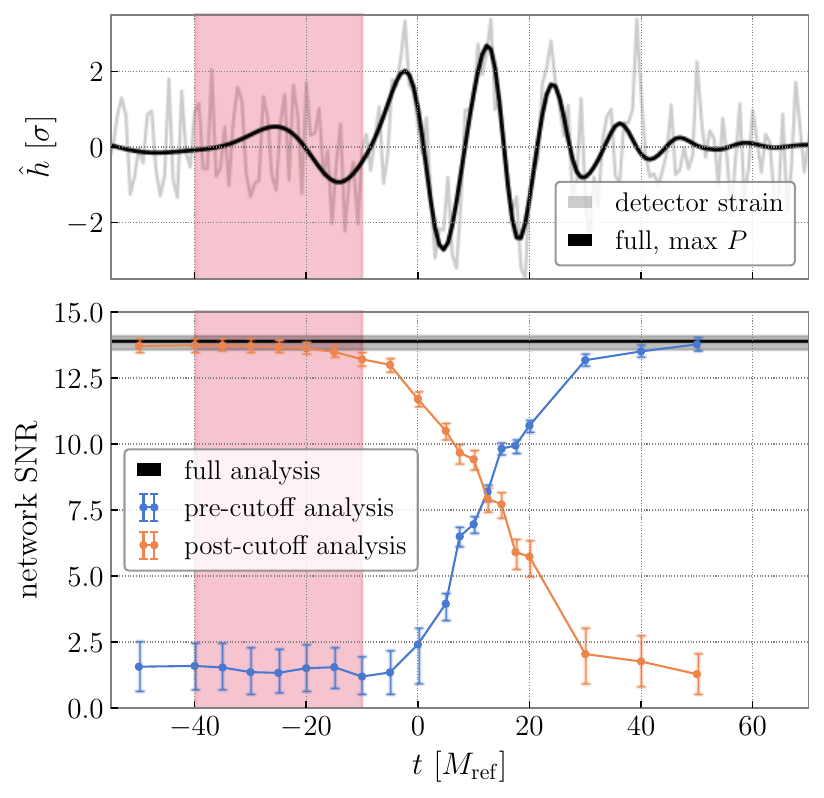}
    \caption{\textit{(Lower subplot)} Network matched-filter SNR for the pre-cutoff (blue) and post-cutoff (orange) analyses as a function of the cutoff time $t$.
    Points indicate the median SNR and error bars represent the 90\% credible region. 
    The horizontal black line represents the median SNR recovered by the full analysis, with the 90\% credible region shaded in gray.
    \textit{(Upper subplot)} Whitened strain data in LIGO Livingston (gray), and whitened waveform reconstruction for the maximum posterior (abbr.~``max $P$") draw from the full analysis (black), plotted to help visualize the signal alongside the SNR evolution.
    In both subplots, the red shaded region -- between $t=-40\,\Mref$ and $-10\,\Mref$ -- is that which is identified as crucial for constraining $\chip$.}
    \label{fig:figure_08}
\end{figure}

\begin{figure*}
    \centering
    \includegraphics[width=0.77\linewidth]{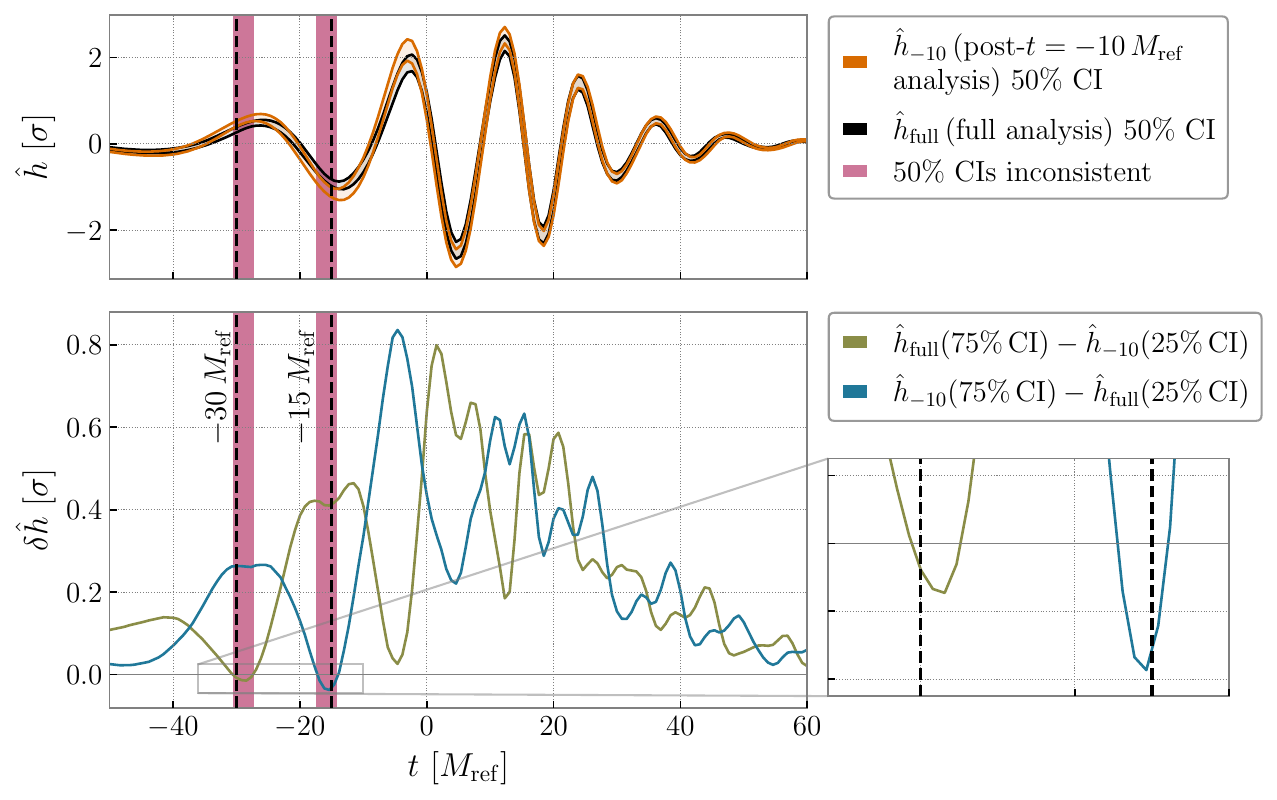}
    \caption{Identification of regions in time (pink shaded) during which the 50\% credible intervals (CI) of the reconstructed waveform strain $\hat h$ are inconsistent between the full (black; labeled $\hat h_\mathrm{full}$) and post-$t=-10\,\Mref$ (orange; labeled $\hat h_\mathrm{-10}$) analyses.
    This occurs around the two extrema of the final pre-merger cycle ($t\sim-30\,\Mref$ and $t\sim-15\,\Mref$; indicated with vertical black dashed lines), when $\delta \hat{h}$---the difference between the 75\%  CI of one analysis and the 25\% CI of the other (green and blue)---is negative (enlarged in the inset).}
    \label{fig:figure_09}
\end{figure*}


\section{Signal strength over time}
\label{app:snr}

We track the time-bounded matched-filter signal-to-noise ratio (SNR) in the LIGO Livingston, LIGO Hanford, and Virgo detector network recovered by our analyses, as defined in Eqs.~(52) and (53) of \citet{Isi:2021iql}.
Figure~\ref{fig:figure_08} shows the SNR evolution for the pre- (blue) and post-cutoff (orange) analyses.
The change in SNR is negligible between $t=-40\,\Mref$ and $-10\,\Mref$ (red shaded region) for both the pre- and post-cutoff analyses, suggesting that the change in inferred $\chip$ between these times is not simply due to a drop in SNR.

To investigate correlations between signal strength and $\chip$ (see Fig.~\ref{fig:figure_02} in the main text), we compare the 50\% credible intervals (CI) of the reconstructed waveform strain for two analyses, one in which precession is inferred (full analysis) and one in which it is not (post-$t=-10\,\Mref$ analysis) . 
The top panel in Fig.~\ref{fig:figure_09} shows the 50\% CIs for the strain $\hat h$ for these two analyses in black and orange respectively.
In the bottom panel, we plot the differences $\delta \hat{h}$ between the 75\% CI of one analysis and the 25\% CI of the other in green and blue.
When either difference is \textit{negative}, the two sets of waveform reconstructions are inconsistent at the 50\% CI.
This inconsistency \textit{only} occurs at the extrema of the final pre-merger cycle, i.e.~the peak around $t\sim-30\,\Mref$ and trough at $t\sim-15\,\Mref$, as indicated by the time-slices shaded in pink. 
These regions are enlarged in the inset.

\bibliographystyle{apsrev4-2} 
\bibliography{OurRefs.bib}

\end{document}